\documentclass[fleqn]{an}

\usepackage{amsmath,graphicx,amssymb}
\usepackage{epsf,psfig}
\usepackage{times}

\begin{document}

\Pagespan{330}{337}
\Yearpublication{2010}
\Yearsubmission{2009}
\Month{11}
\Volume{331}
\Issue{3}
\DOI{10.1002/asna.200911302}

\sloppy
\overfullrule5pt

\title{The evolution of chaos in active galaxy models with an oblate or a \\
 prolate dark halo component}

\author{Nicolaos D. Caranicolas\inst{}
\and Euaggelos E. Zotos\inst{}\fnmsep\thanks{Corresponding author:
\email{evzotos@astro.auth.gr}}}

\titlerunning{Evolution of chaos in active galaxies with a dark halo component}
\authorrunning{N. D. Caranicolas \& E. E. Zotos}

\institute{
Department of Physics, Section of Astrophysics, Astronomy and Mechanics, Aristotle University of Thessaloniki, \\
GR-541 24, Thessaloniki, Greece}

\received{2009 Jun 10}
\accepted{2009 Nov 6}
\publonline{2010 Mar 12}


\abstract{The evolution of chaotic motion in a galactic dynamical model with a disk, a dense nucleus and a flat biaxial dark halo component is investigated. Two cases are studied: (i) the case where the halo component is oblate and (ii) the case where a prolate halo is present. In both cases, numerical calculations show that the extent of the chaotic regions decreases
exponentially as the scale-length of the dark halo increases. On the other hand, a linear relationship exists between the extent of the chaotic regions and the flatness parameter of the halo component. A linear relationship between the critical
value of the angular momentum and the flatness parameter is also found. Some theoretical arguments to support the numerical outcomes are presented. An estimation of the degree of chaos is made by computing the Lyapunov Characteristic Exponents. Comparison with earlier work is also made.}

\maketitle

\section{Introduction}

Observational data show that disk galaxies are often surrounded by massive and extended dark matter haloes. The best tool to study dark matter haloes in galaxies are the rotation curves derived from neutral hydrogen (see e.g. Meurer et al. 1996; Dinshaw et al. 1998; Shull et al. 1998; Oppenheimer et al. 2001; McLin et al. 2002; Penton et al. 2002; Steidel et al. 2002; Cote et al. 2005). On the other hand, numerical simulations suggest that dark matter galactic haloes are not only spherical but also may be oblate, prolate or triaxial (Merritt \& Fridman 1996; Cooray 2000; Kunihito et al. 2000; Olling \& Merrifield 2000;  Jing \& Suto 2002; Wechsler et al. 2002; Kasun \& Evrard 2005; Allgood et al. 2006; Capuzzo-Dolcetta et al. 2007; Wang et al. 2009; Evans et al. 2009; Caranicolas \& Zotos 2009). The variety of the shapes of galactic haloes strongly indicates that the structure of these objects plays an important role in the orbital behavior and, generally, in the dynamics of a galaxy.

In two earlier papers (Caranicolas 1997; Papadopoulos \& Caranicolas 2006) we have studied axially symmetric or non axially symmetric active galaxy models with an additional spherical halo component. In both cases, it was observed that the presence of the spherical halo had as a result to reduce the area in the phase space occupied by the chaotic orbits. Therefore, it would be of significant interest to investigate the behavior of orbits in an active galaxy with a biaxial halo. On this basis, we have decided to study the motion in a disk galaxy model with an oblate or a prolate dark halo component. Particular interest will be given to the study of the regular or chaotic character of orbits and its connection to the physical parameters of the system,
such as the flatness, the core radius of the halo component and the conserved component of the angular momentum. Furthermore, we shall use the Lyapunov Characteristic Exponent (L.C.E) (see Lichtenberg \& Lieberman 1992) in order to estimate and compare the degree of chaos in each case.

The dynamical model is presented in Sect. 2. Two cases are distinguished. The case when an oblate and the case when a prolate halo component is present. In Sect. 3 we study the behavior of orbits when an oblate or a prolate halo component is present. In the same section, numerically found relationships between the extent of the chaotic regions and the physical parameters of the system are presented. In Sect. 4 some semi-theoretical arguments are used in order to explain the numerically obtained results. We close with Sect. 5 where a discussion and the conclusions of this research are given and a comparison with previous work is made.

\section{Description of the dynamical model}

There are three components in our dynamical model. The disk-halo, the dense nucleus and the dark halo component. The disk-halo component is represented by the potential
\begin{flalign}
&V_d(r,z) = \frac{-M_d}{R},&
\end{flalign}
where $R^2 = b^2 + r^2 + \left(\alpha + \sqrt{h^2 + z^2}\right)^2$. Here $(r,z)$ are the usual cylindrical coordinates, $M_d$ is the mass, $b$ is the core radius of the disk-halo, $\alpha$ is the disk's scale length and $h$ corresponds to the disk's scale height. The dense nucleus is represented by the spherical potential
\begin{flalign}
&V_n(r,z) = \frac{-M_n}{\left(r^2 + z^2 + c_n^2\right)^{1/2}},&
\end{flalign}
where $M_n$ is the mass and $c_n$ is the scale length of the nucleus. For the dark halo component we use the logarithmic potential
\begin{flalign}
&V_h(r,z) = \frac{\upsilon_0^2}{2} \ln \left(r^2 + \beta z^2 + c_h^2 \right),&
\end{flalign}
where $\beta$ is the flatness parameter, while $c_h$ stands for the scale length of the dark halo component. The parameter $\upsilon_0$ is used for the consistency of the galactic units. We have chosen this potential because Eqs. (1) and (2) represent a modified version of the analytical potential used successfully by Caranicolas \& Innanen (1991) to describe the motion in an active disk galaxy. Furthermore, Eq. (3) is used because we believe that potential (3) is suitable for the description of the motion in a dark halo as it produces a flat rotation curve (see Binney \& Tremaine 2008). In this research,
we shall use a system of galactic units, where the unit of length is 1 kpc, the unit of time is 0.977 $\times$ $10^8$ yr and the unit of mass is 2.325 $\times$ $10^7$ $M_{\odot}$. The velocity unit is 10 km/s, while $G$ is equal to unity. In the above units we use the values: $\upsilon_0 = 20, \alpha = 3, b = 6, h = 0.2, M_d = 5000, M_n = 400, c_n = 0.25$, while $\beta$ and $c_h$  are treated as parameters. The total potential responsible for the motion of a test particle (star) of unit mass in the galaxy is
\begin{flalign}
&V_t(r,z) = V_d + V_n + V_h.&
\end{flalign}
As the total potential $V_t = V_t(r,z)$ is axially symmetric and the $L_z$ component of the angular momentum is conserved we use the effective potential
\begin{flalign}
&V_{eff}(r,z) = \frac{L_z^2}{2r^2} + V_t(r,z),&
\end{flalign}
in order to study the motion in the meridian $(r,z)$ plane. The equations of motion are
\begin{flalign}
&\dot{r} = p_r, \ \ \ \dot{z} = p_z, \ \ \
\dot{p_r} = -\frac{\partial V_{eff}}{\partial r}, \ \ \
\dot{p_z} = -\frac{\partial V_{eff}}{\partial z},&
\end{flalign}
where the dot indicates derivative with respect to the time. The corresponding Hamiltonian is written as
\begin{flalign}
&H = \frac{1}{2}(p_r^2 + p_z^2) + V_{eff}(r,z) = E,&
\end{flalign}
where $p_r$ and $p_z$ are the momenta, per unit mass, conjugate to $r$ and $z$, while $E$ is the numerical value of the Hamiltonian, which is conserved. Eq. (7) is an integral of motion, which indicates that the total energy of the test particle is conserved.

Orbit calculations are based on the numerical integration of the equations of motion, which was made using a sharp Bulirsh-St\"{o}er FORTRAN routine in double precision. By the term sharp we mean that all subroutines of the numerical integration code are in double precision. The accuracy of the calculations was checked by the constancy of the energy integral, which was conserved up to the twelfth significant figure.

\section{Orbit calculations when a biaxial halo component is present}

In this section we shall study the behavior of orbits when a biaxial halo is present. We shall use the classical method of the Poincar\'{e} $(r,p_r)$, $z=0, p_z > 0$ phase plane in order to determine the regular or chaotic character of motion. Two cases will be studied: (a) the system has a spherical or oblate halo, that is when $1 \leq \beta < 2$ and (b) the system has a prolate halo, that is when $0.1 \leq \beta <1$. The value of $c_h$ is in the range $8.5 \leq c_h \leq 21$.

\subsection{Model with an oblate halo}

Figure 1a-b shows the $(r,p_r)$ phase plane when (a) $\beta = 1.3$ and (b) $\beta = 1.8$. The value of $c_h$ is 8.5, while the values of all the other parameters are $\upsilon_0 = 20, \alpha = 3, b = 6, h = 0.2, M_d = 5000, M_n = 400, c_n = 0.25$ and $L_z = 10$. Here, we must note that all the initial conditions are taken inside the limiting curve. This curve contains all the invariant curves in the $(r,p_r)$, $z=0, p_z>0$  phase plane and can be obtained from (7) by setting $z = p_z = 0$ (see Papadopoulos \& Caranicolas 2006). Thus, we take the values of $r$ and $p_r$ inside the limiting curve, while the value of $p_z$ is found from the energy integral (7). As one can see the majority of the phase plane is covered by chaotic orbits. Regular orbits are confined mainly near the central parts of the phase plane. There are also some smaller islands produced by secondary resonances. One observes that in both cases there is a single chaotic sea. This chaotic sea is larger in the case  when $\beta = 1.8$. This suggests that the flatness parameter of the halo plays an important role on the character of motion. Our numerical calculations indicate that the chaotic region in disk galaxies with dense nuclei increases when the flatness parameter $\beta$ of the halo increases, provided that all the other parameters are kept constant. We shall come to this
point later in the next section.
\begin{figure*}
\resizebox{\hsize}{!}{\rotatebox{0}{\includegraphics*{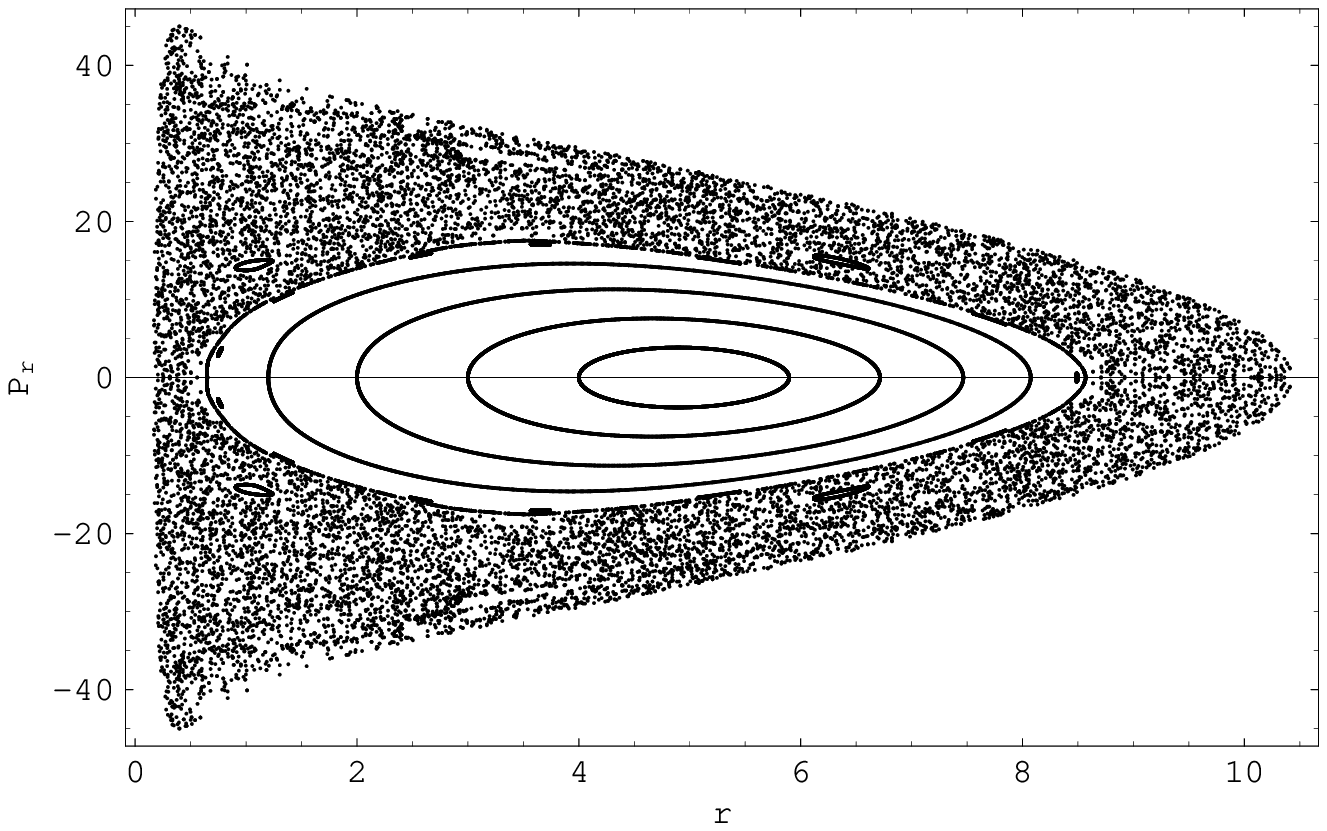}}\hspace{1cm}
                      \rotatebox{0}{\includegraphics*{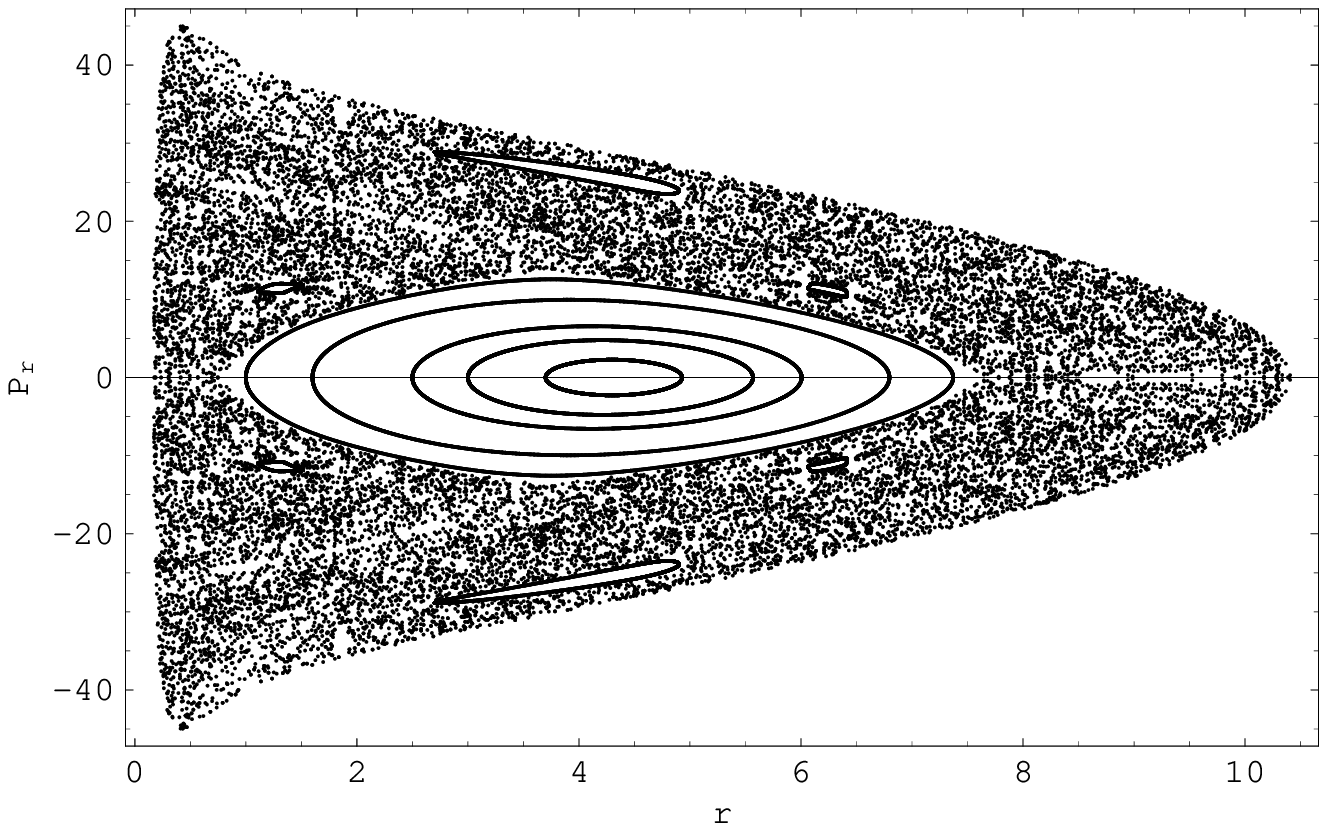}}}
\vskip 0.1cm
\caption{(a-b): The $(r,p_r)$ phase plane when (a, \textit{left}) $\beta = 1.3$ and (b, \textit{right}) $\beta = 1.8$. The value of $c_h$ is 8.5, while the values of all the other parameters are: $\upsilon_0 = 20, \alpha = 3, b = 6, h = 0.2, M_d = 5000, M_n = 400, c_n = 0.25$ and $L_z = 10.$}
\end{figure*}

Figure 2a-b shows the $(r,p_r)$ phase plane when (a) $c_h = 11$ and (b) $c_h = 18.5$. The value of $\beta$ is 1.7, while all the other parameters are as in Fig. 1. Again, the majority of orbits are chaotic in both cases. Here, we see that when $c_h = 11$ the chaotic sea is larger. This indicates that the scale length of the halo plays also an important role on the character of motion. We observe that the extent of the chaotic regions in the phase plane decreases as $c_h$ increases, when all the other parameters are kept constant. Therefore, we conclude that the chaotic regions in active disk galaxies are larger when a dense halo component is present. Figure 3 shows the relationship between the percentage of the area $A\%$ covered by chaotic orbits in the phase plane and $c_h$ when $\beta = 1.7$. We see that $A\%$ decreases exponentially as $c_h$ increases. Here, we must notice that the whole area of the phase plane is reduced and becomes smaller as $c_h$ increases.
\begin{figure*}
\resizebox{\hsize}{!}{\rotatebox{0}{\includegraphics*{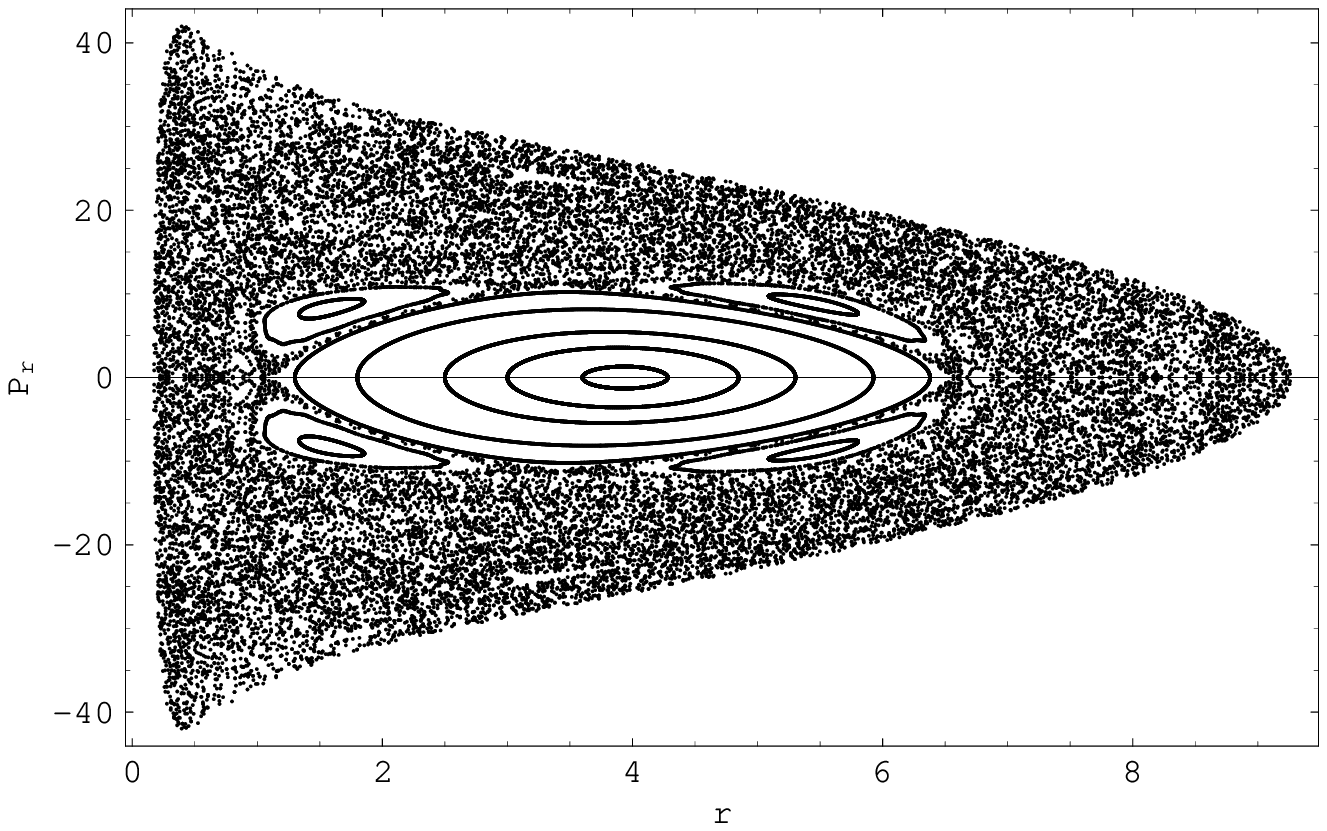}}\hspace{1cm}
                      \rotatebox{0}{\includegraphics*{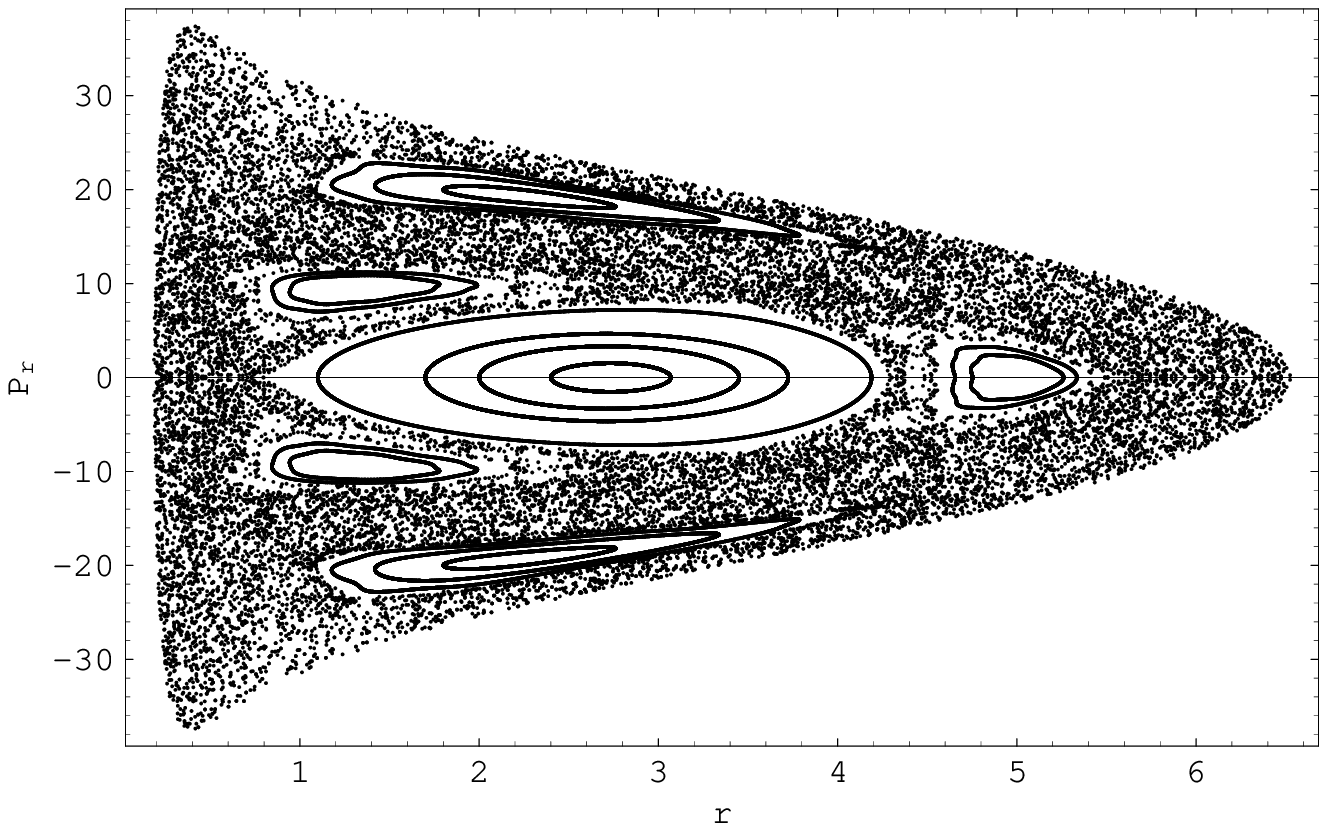}}}
\vskip 0.1cm
\caption{(a-b): The $(r,p_r)$ phase plane when (a, \textit{left}) $c_h = 11$ and (b, \textit{right}) $c_h = 18.5$. The value of $\beta$ is 1.7, while the values of all the other parameters are as in Fig. 1.}
\end{figure*}
\begin{figure}
\resizebox{\hsize}{!}{\rotatebox{0}{\includegraphics*{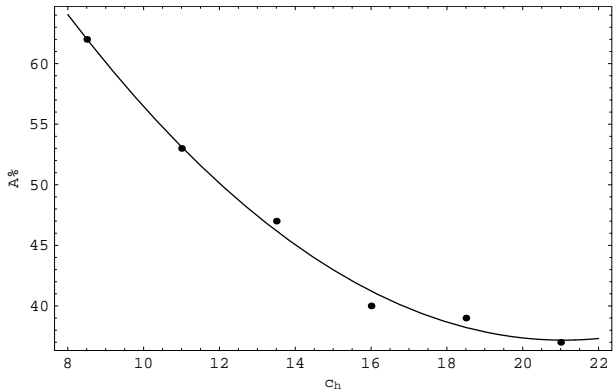}}}
\caption{A plot of the percentage of the area $A\%$ in the phase plane covered by chaotic orbits vs. $c_h$ when $\beta = 1.7$}
\end{figure}

Figure 4a-d shows four representative orbits. Figure 4a shows a regular orbit when $\beta = 1$ and $c_h=8.5$. The initial conditions are: $r_0 = 3, z_0 = 0, p_{r0} = 30$. The value of $p_{z0}$ is found from the energy integral (7) for all orbits. Figure 4b shows a regular orbit orbit when $\beta = 1.7$ and $c_h=13$. The initial conditions are: $r_0 = 6, z_0 = 0, p_{r0} = 0$. In Fig. 4c a quasi periodic orbit is shown. Here, $\beta = 1.9$ and $c_h=8.5$. The initial conditions are: $r_0 = 9, z_0 = 0, p_{r0} = 0$. This orbit is characteristic of the 4:3 resonance. A chaotic orbit is given in Fig. 4d. The value of $\beta$ is 1.6, while $c_h=8.5$. The initial conditions are: $r_0 = 5, z_0 = 0, p_{r0} = 20$. All orbits were calculated for a time period of 100 time units. The value of energy is $E = 600$, while the values of all the other parameters are: $\upsilon_0 = 20, \alpha = 3, \beta = 6, h = 0.2, M_d = 5000, M_n = 400$ and $c_n = 0.25$. The value of $L_z$ is equal to 10.
\begin{figure*}
\resizebox{\hsize}{!}{\rotatebox{0}{\includegraphics*{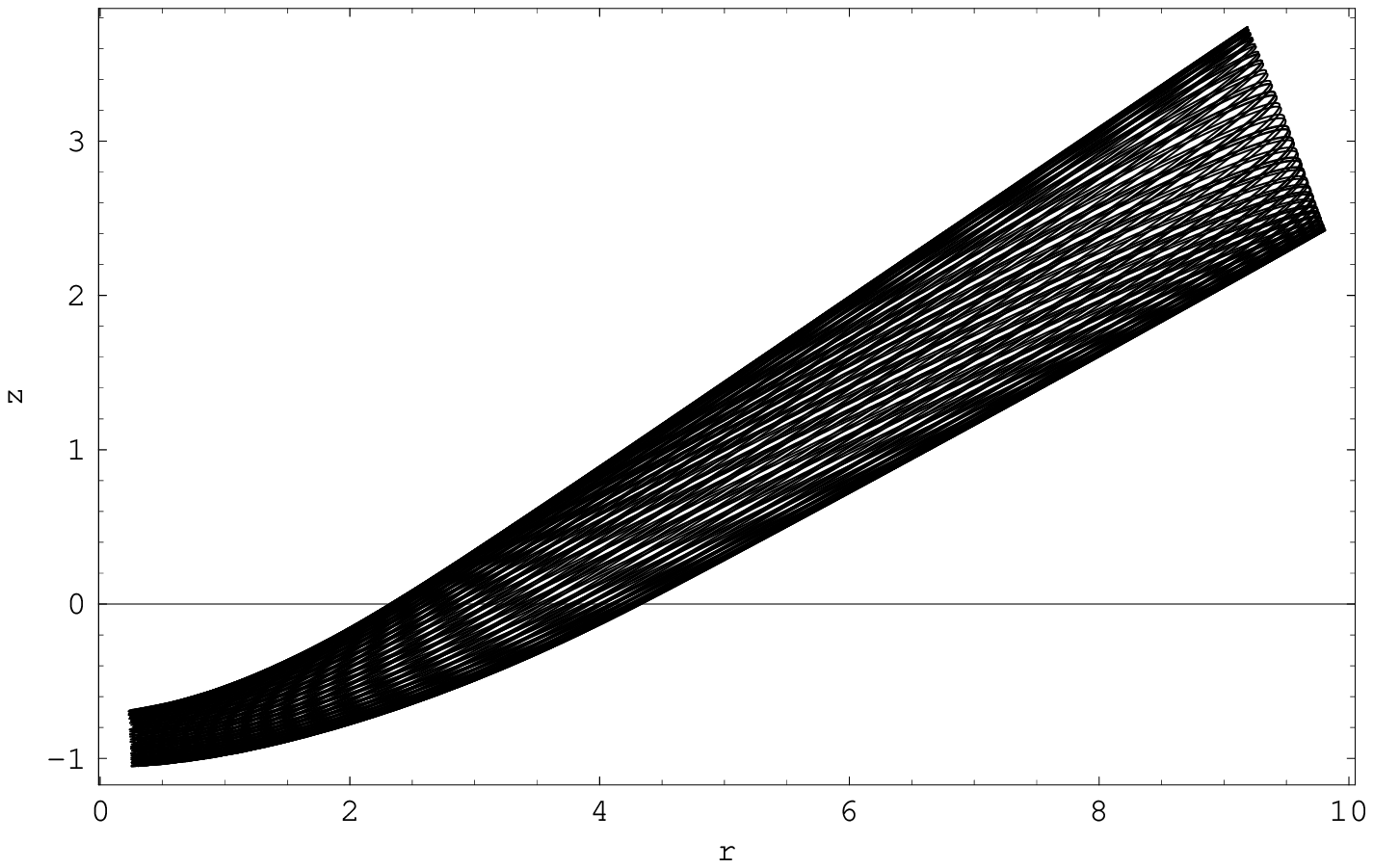}}\hspace{1cm}
                      \rotatebox{0}{\includegraphics*{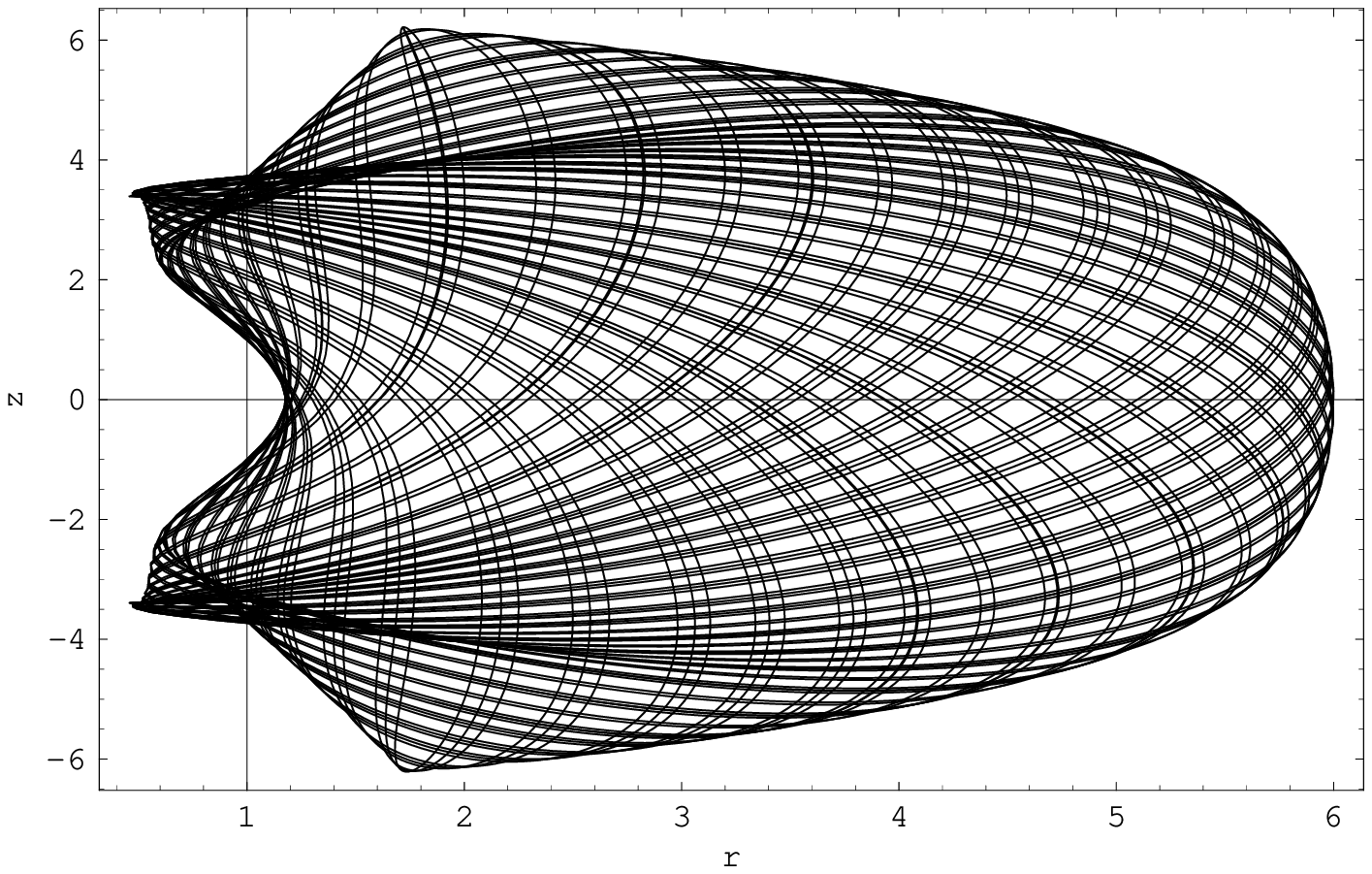}}}
\vskip 0.1cm
\resizebox{\hsize}{!}{\rotatebox{0}{\includegraphics*{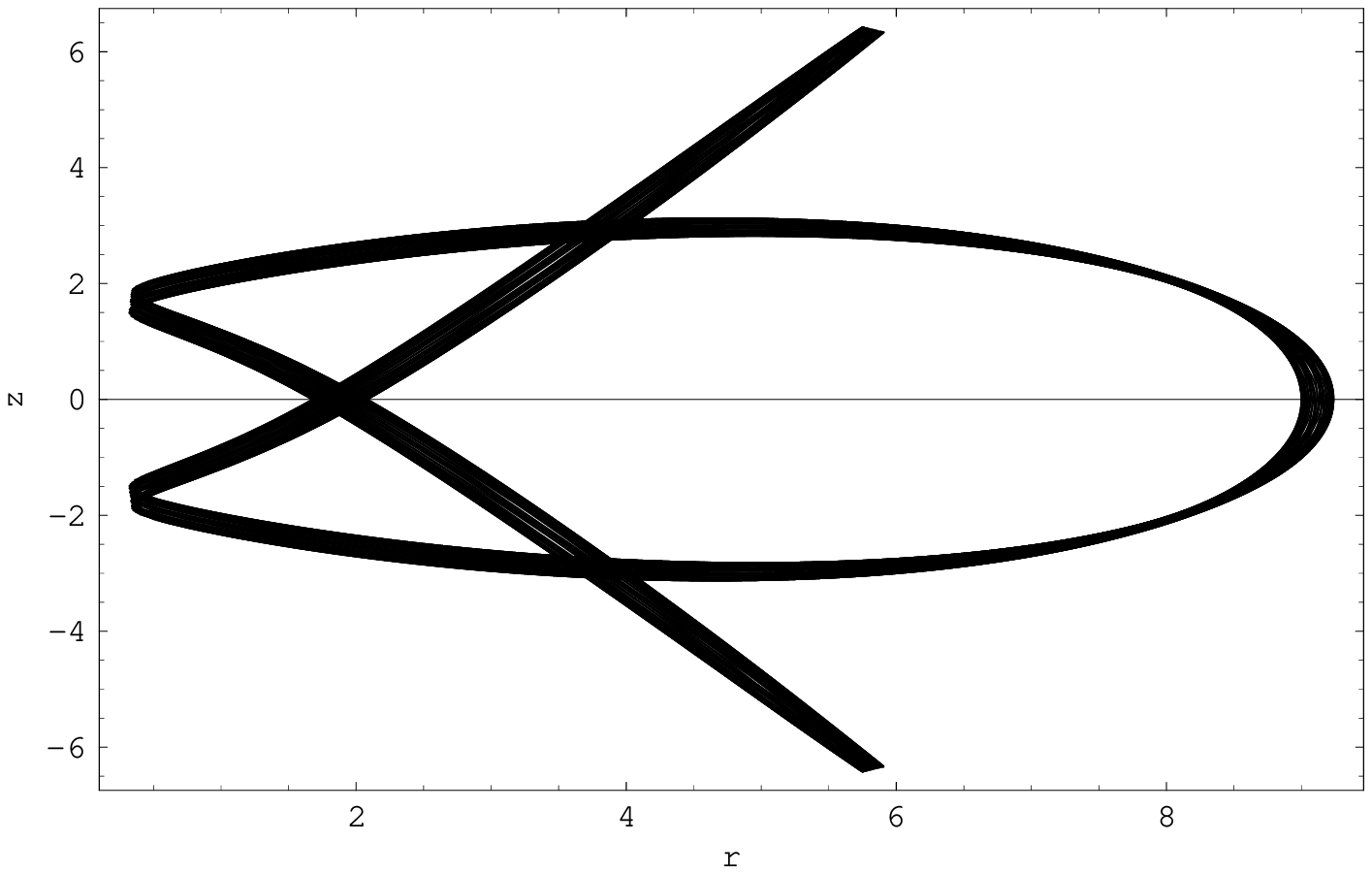}}\hspace{1cm}
                      \rotatebox{0}{\includegraphics*{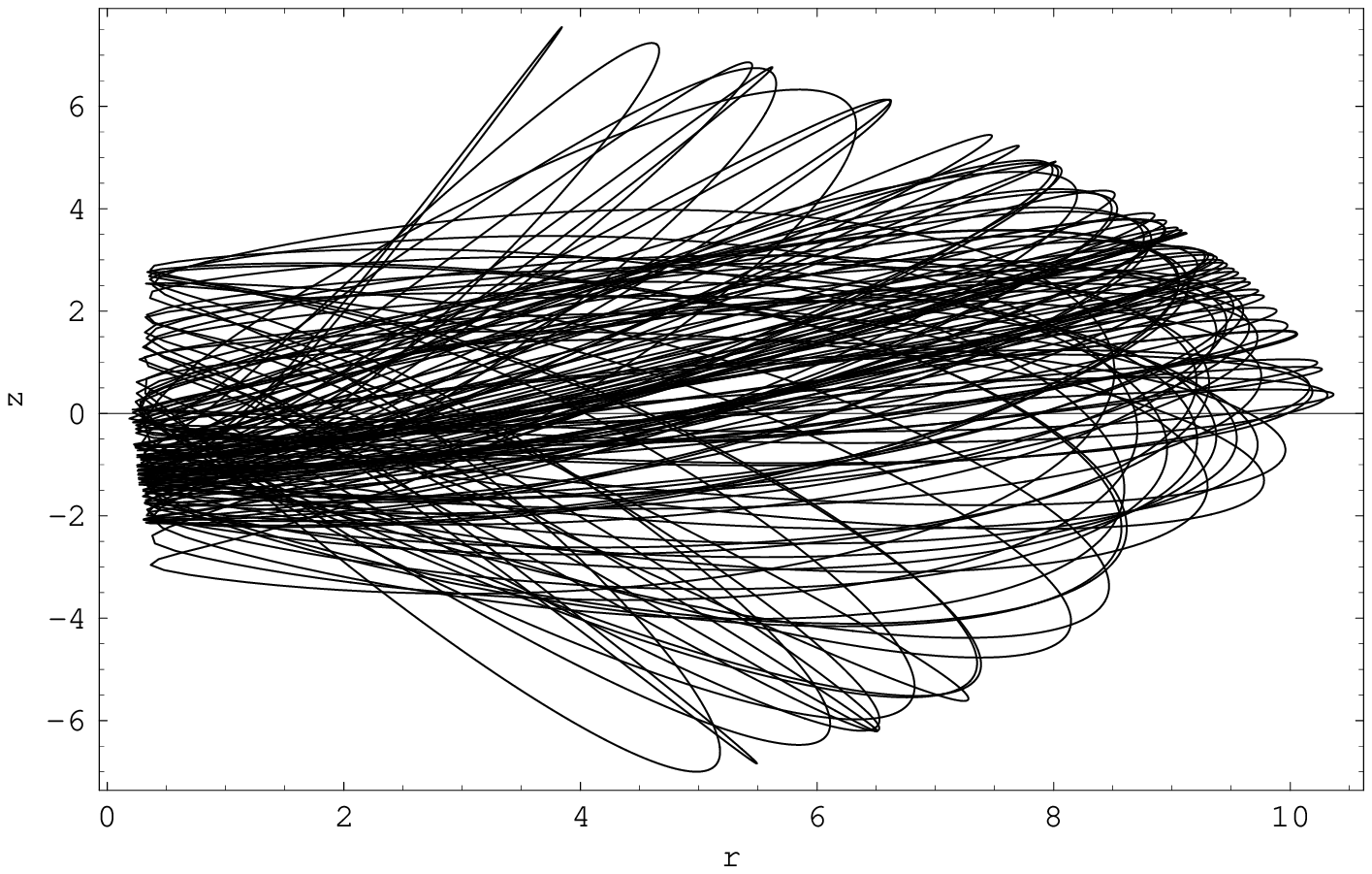}}}
\vskip 0.1cm
\caption{(a-d): Orbits when the system has an oblate dark halo. (a, \textit{upper left}): a regular orbit when $\beta = 1.0$ and $c_h=8.5$. The initial conditions are: $r_0 = 3, z_0 = 0, p_{r0} = 30$. (b, \textit{upper right}): a regular orbit orbit when $\beta = 1.7$ and $c_h=13$. The initial conditions are: $r_0 = 6, z_0 = 0, p_{r0} = 0$. (c, \textit{lower left}): a quasi periodic orbit when $\beta = 1.9$ and $c_h = 8.5$. The initial conditions are: $r_0 = 9, z_0 = 0, p_{r0} = 0$. (d, \textit{lower right}): a chaotic orbit. The value of $\beta$ is 1.6, while $c_h = 8.5$. The initial conditions are: $r_0 = 5, z_0 = 0, p_{r0} = 20$. The value of $p_{z0}$ is found from the energy integral (7) for all orbits. The values of all the other parameters are given in text.}
\end{figure*}

In order to have an estimation of the degree of chaos from a different point of view, we computed the L.C.E in the chaotic sea for each of the corresponding $(r,p_r)$ phase planes for a time period of $10^5$ time units for different values of $\beta$ and $c_h$. Our numerical experiments suggest that the L.C.E is in the range from about 0.36 to about 0.43, when $c_h = 8.5$ and
$1 \leq \beta < 2$, while it is in the range from about 0.40 to about 0.55, when $\beta = 1.7$ and $8.5 \leq c_h \leq 21$. Figure 5 shows the L.C.E for the chaotic orbit of Fig. 4d.
\begin{figure}
\resizebox{\hsize}{!}{\rotatebox{0}{\includegraphics*{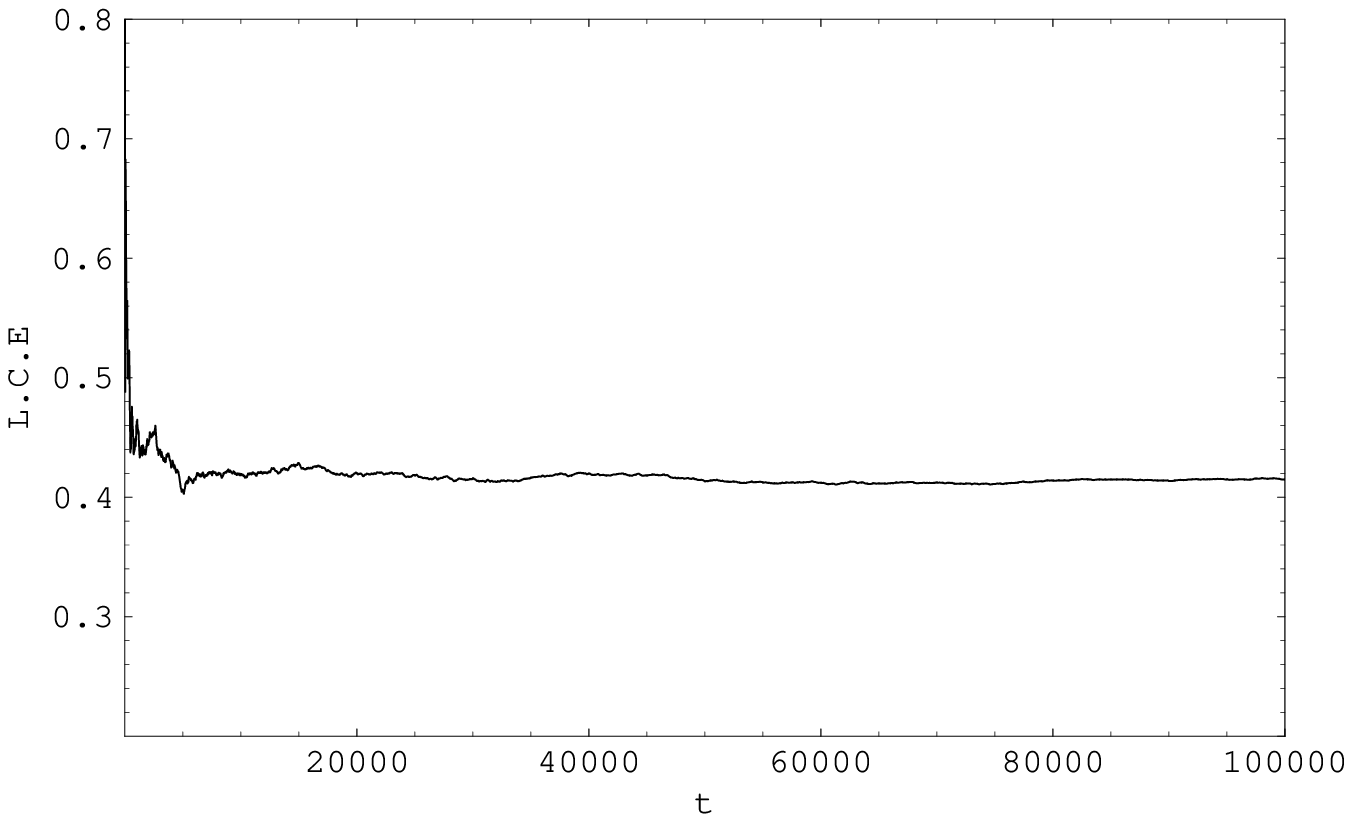}}}
\caption{Evolution of the L.C.E with the time for the chaotic orbit shown in Fig. 4d.}
\end{figure}

\subsection{Model with a prolate halo}

Figure 6a-b shows the $(r,p_r)$ phase plane when (a) $\beta = 0.6$ and (b) $\beta = 0.2$. The value of $c_h$ is $8.5$, while the values of all the other parameters are: $\upsilon_0 = 20, \alpha = 3, \beta = 6, h = 0.2, M_d = 5000, M_n = 400, c_n = 0.25$ and $L_z=10$. Let us start from Fig. 6a. In this case we see that the majority of the phase plane is covered by regular orbits. A considerable chaotic layer is present and it is confined in the outer parts of the phase plane. In Fig. 6b we have the case where $\beta = 0.2$. Here again the majority of the orbits are regular and secondary resonances are also present. The main difference from Fig. 6a is that now the chaotic layer has become a chaotic sea. This suggests that the flatness parameter of the prolate halo plays also an important role on the character of motion. We shall come to this point later in the next section.
\begin{figure*}
\resizebox{\hsize}{!}{\rotatebox{0}{\includegraphics*{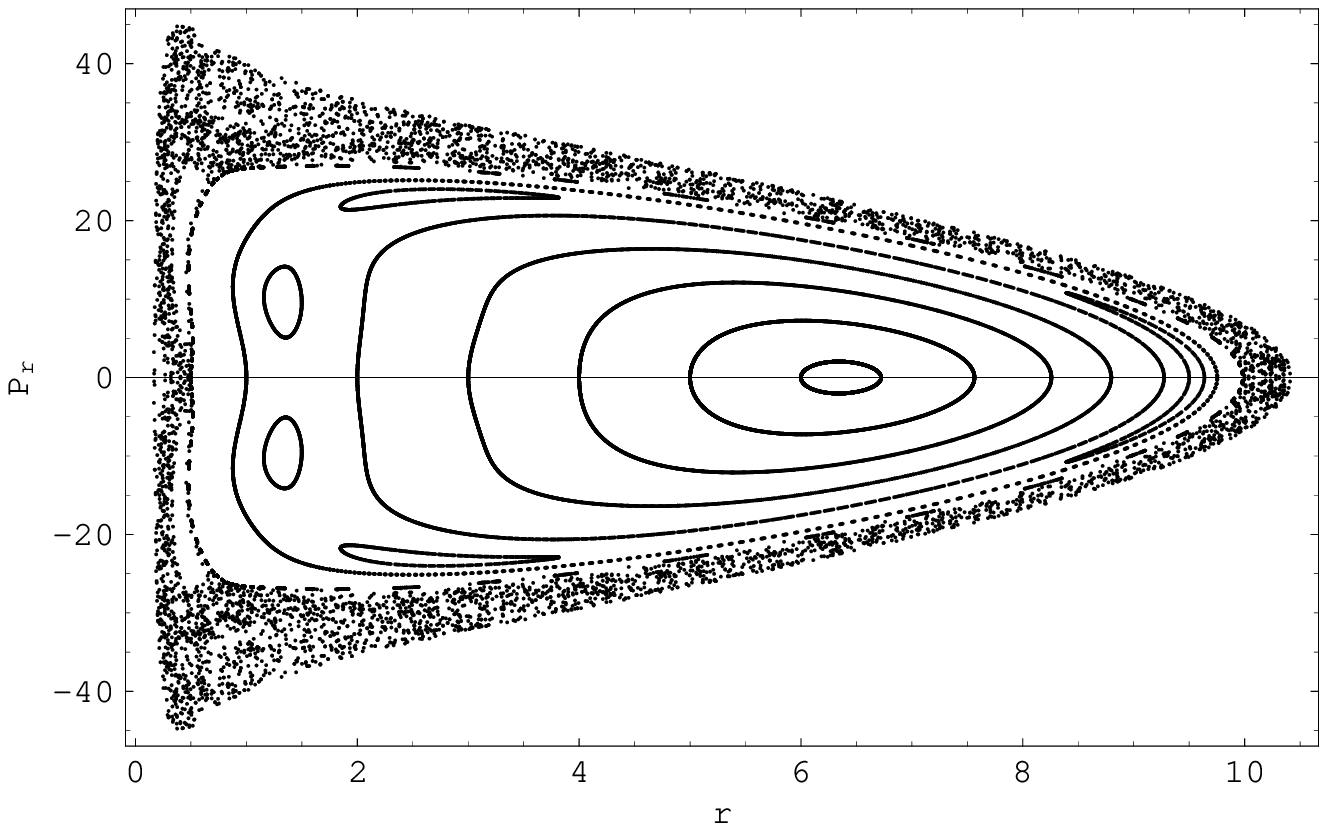}}\hspace{1cm}
                      \rotatebox{0}{\includegraphics*{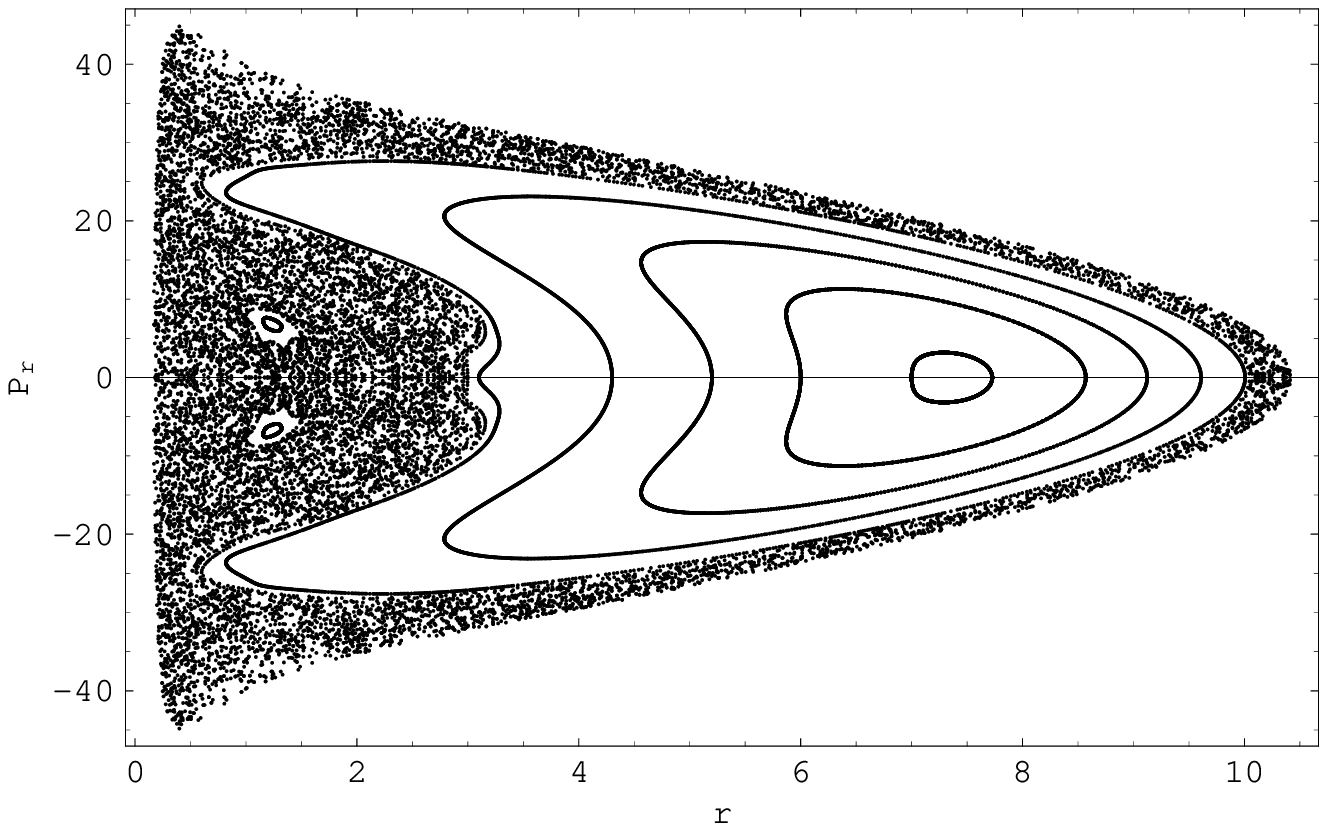}}}
\vskip 0.1cm
\caption{(a-b): The $(r,p_r)$ phase plane when (a, \textit{left}) $\beta = 0.6$ and (b, \textit{right}) $\beta = 0.2$. The value of $c_h$ is 8.5, while the values of all the other parameters are: $\upsilon_0 = 20, \alpha = 3, \beta = 6, h = 0.2, M_d = 5000, M_n = 400, c_n = 0.25$ and $L_z = 10$.}
\end{figure*}

Figure 7a-b shows the $(r,p_r)$ phase plane when (a) $c_h = 11$ and (b) $c_h = 18.5$. The value of $\beta$ is 0.6, while all the other parameters are as in Fig. 1. Again, the majority of orbits are regular. A careful observation shows that when $c_h = 11$ the chaotic sea is larger than when $c_h = 18.5$. Note again that the whole area of the phase plane is reduced and becomes smaller as $c_h$ increases. Figure 8 shows the relationship between the percentage of the area $A\%$ covered by chaotic orbits in the phase plane and $c_h$, when $\beta = 0.7$. We observe that $A\%$ decreases exponentially as $c_h$ increases. This behavior is similar to that found in Fig. 3 for the oblate halo component.
\begin{figure*}
\resizebox{\hsize}{!}{\rotatebox{0}{\includegraphics*{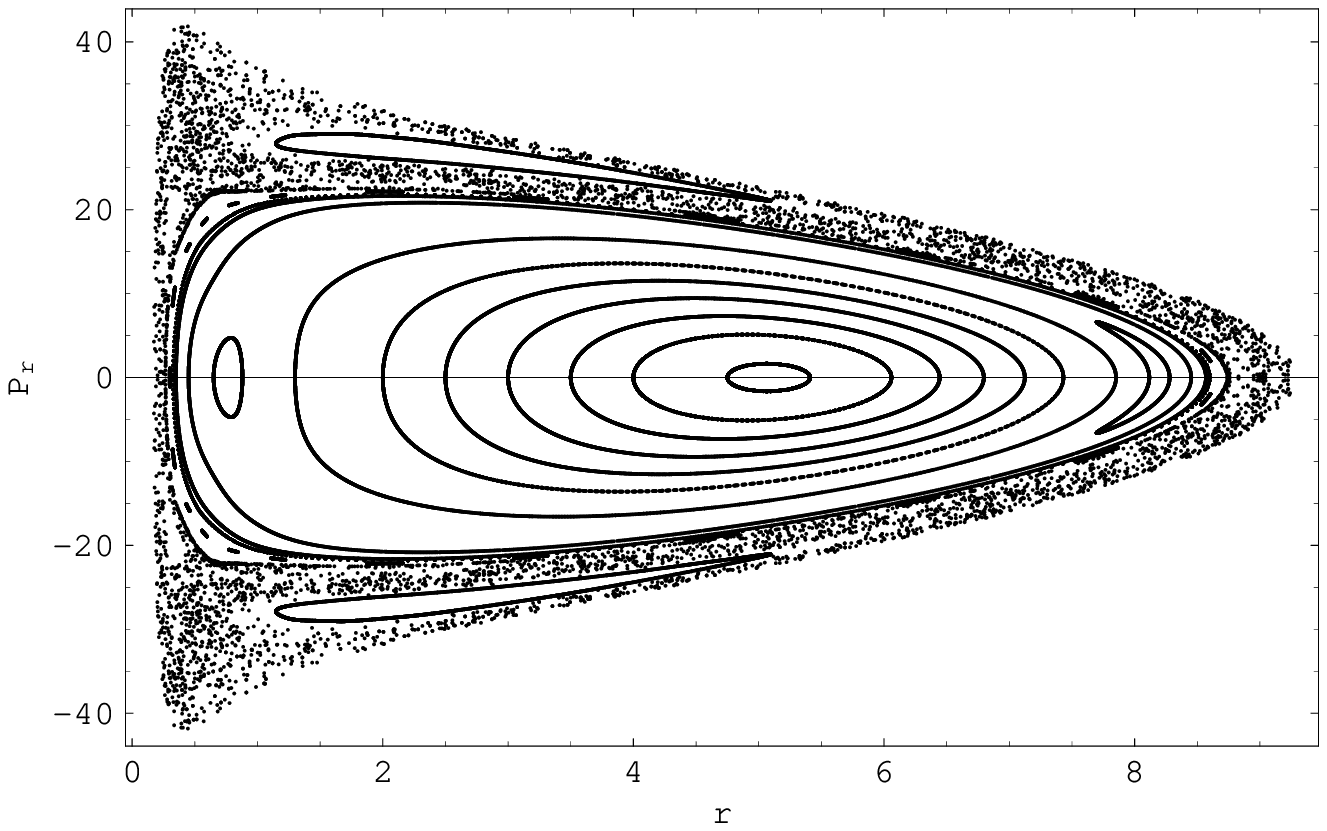}}\hspace{1cm}
                      \rotatebox{0}{\includegraphics*{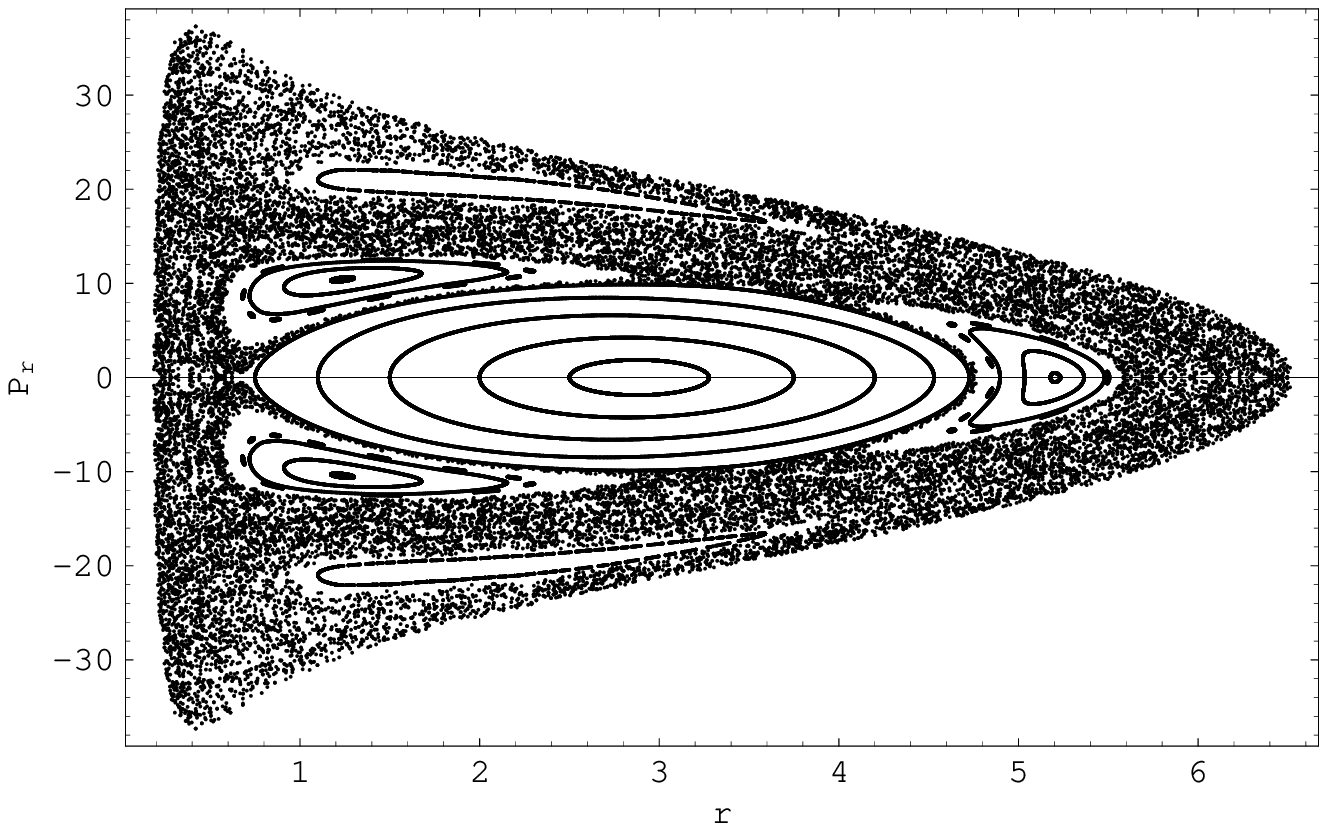}}}
\vskip 0.1cm
\caption{(a-b): The $(r,p_r)$ phase plane when (a, \textit{left}) $c_h = 11$ and (b, \textit{right}) $c_h = 18.5$. The value of $\beta$ is 0.6, while the values of all the other parameters are as in Fig. 6.}
\end{figure*}
\begin{figure}
\resizebox{\hsize}{!}{\rotatebox{0}{\includegraphics*{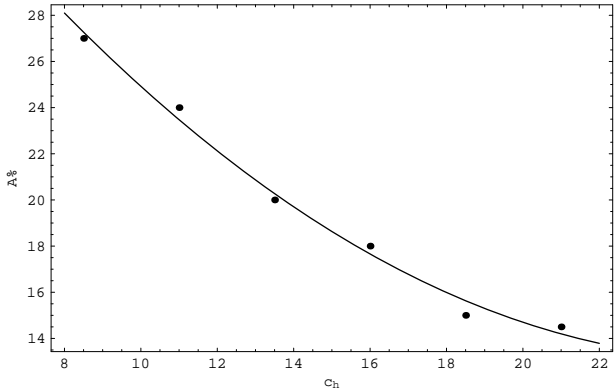}}}
\caption{A plot of the percentage of the area $A\%$ in the phase plane covered by chaotic orbits vs. $c_h$ when $\beta = 0.7$. The values of all the other parameters are given in text.}
\end{figure}

In Figure 9a-d we present four representative orbits. Figure 9a shows an orbit when $\beta = 0.2$ and $c_h = 8.5$. The initial conditions are: $r_0 = 1.3, z_0 = 0, p_{r0} = 6$. This orbit produces the upper small island shown in Fig. 5b. Figure 6b shows a regular orbit when $\beta = 0.3$ and $c_h = 8.5$. The initial conditions are: $r_0 = 5, z_0 = 0, p_{r0} = 0$. The orbit is characteristic of the 2:1 resonance. In Fig. 9c a quasi periodic orbit is shown. Here, $\beta = 0.6$ and $c_h = 16$. The initial conditions are: $r_0 = 6.5, z_0 = 0, p_{r0} = 0$. The orbit is characteristic of the 4:3 resonance. A chaotic orbit is given in Fig. 9d. The value of $\beta$ is 0.2, while $c_h = 8.5$. The initial conditions are: $r_0 = 2, z_0 = 0, p_{r0} = 0$. All orbits were calculated for a time period of 100 time units. The value of energy is $E = 600$, while the values of all the other parameters are: $\upsilon_0 = 20, \alpha = 3, \beta = 6, h = 0.2, M_d = 5000, M_n = 400, c_n = 0.25$ and $L_z = 10$.
\begin{figure*}
\resizebox{\hsize}{!}{\rotatebox{0}{\includegraphics*{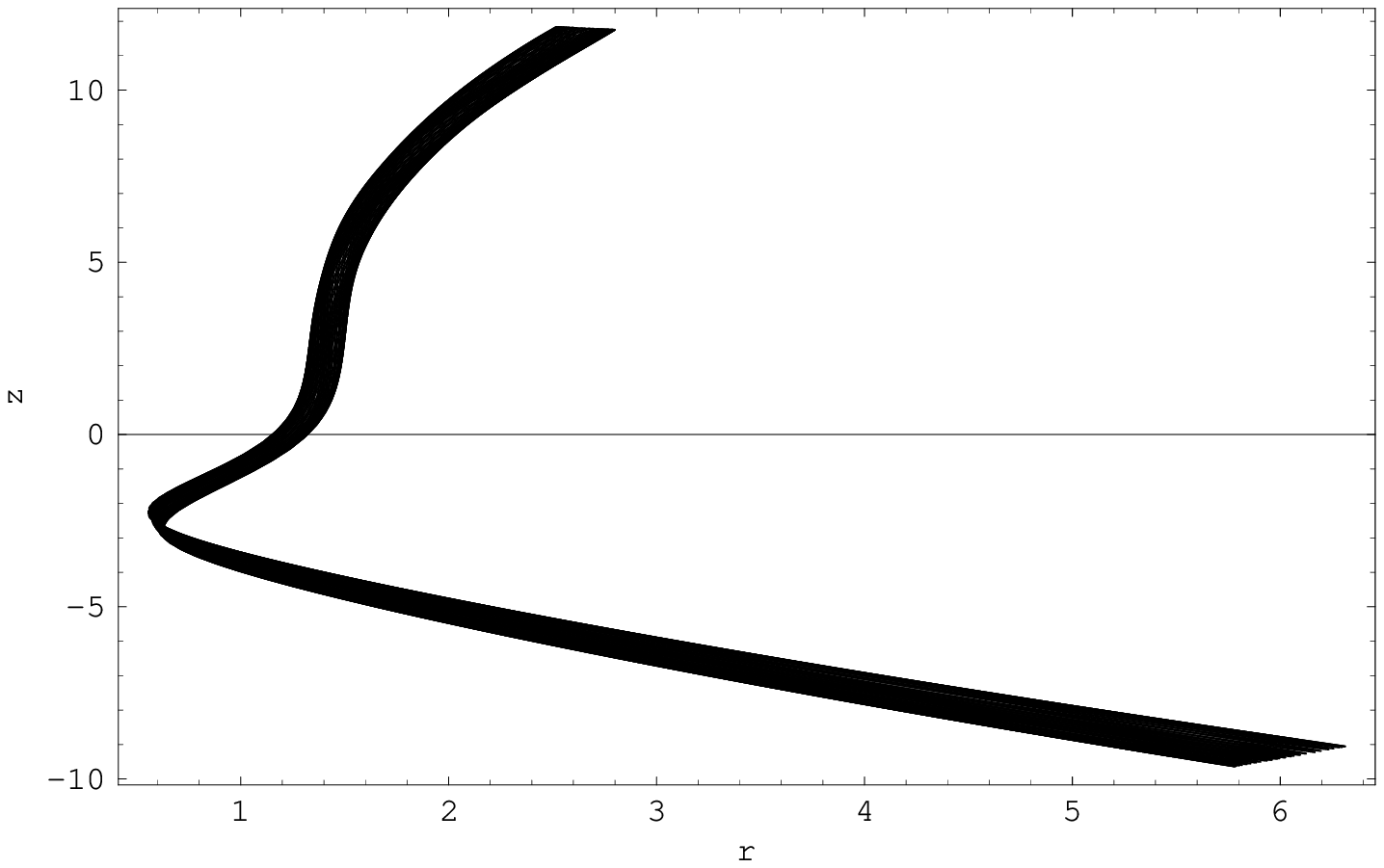}}\hspace{1cm}
                      \rotatebox{0}{\includegraphics*{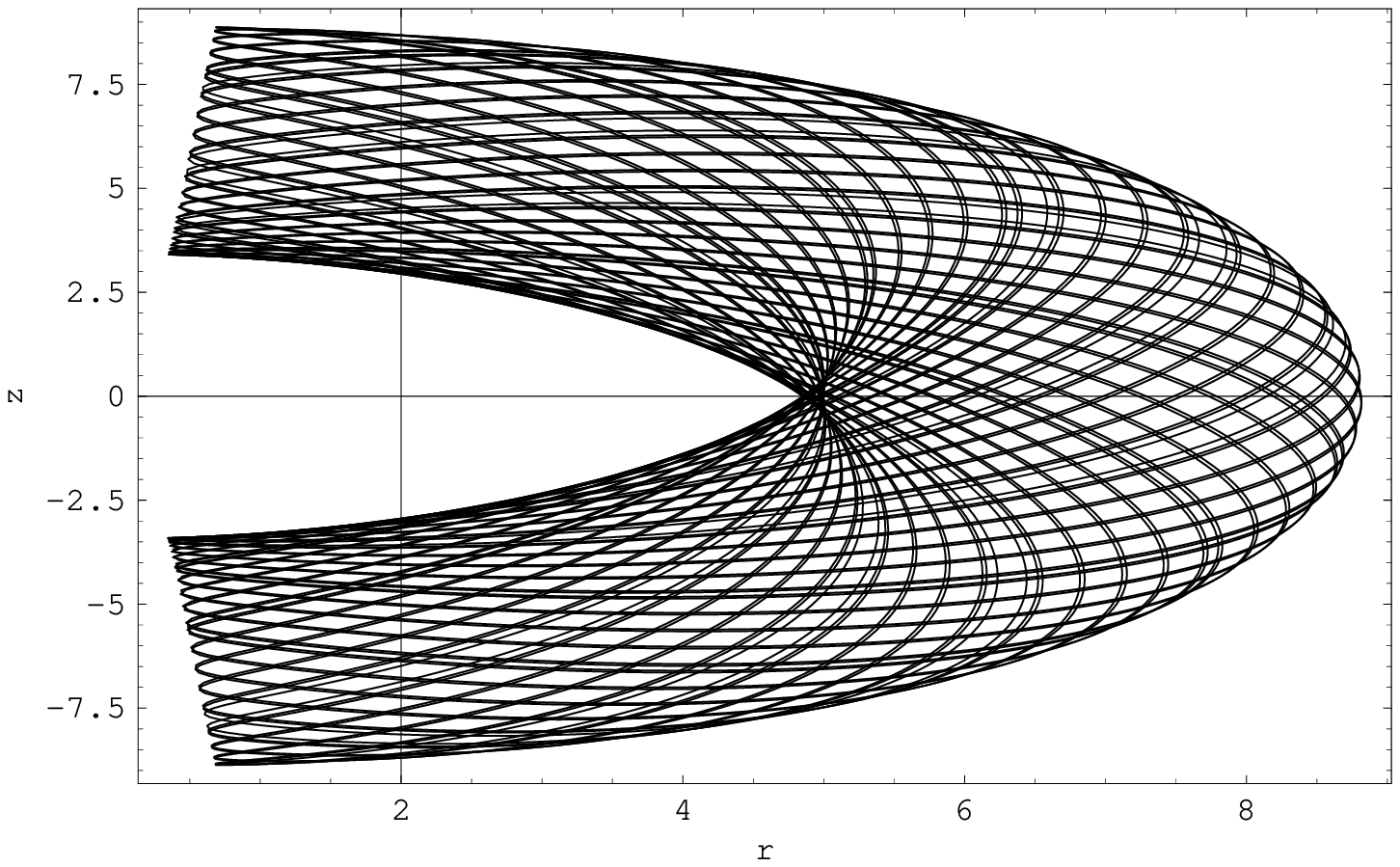}}}
\vskip 0.1cm
\resizebox{\hsize}{!}{\rotatebox{0}{\includegraphics*{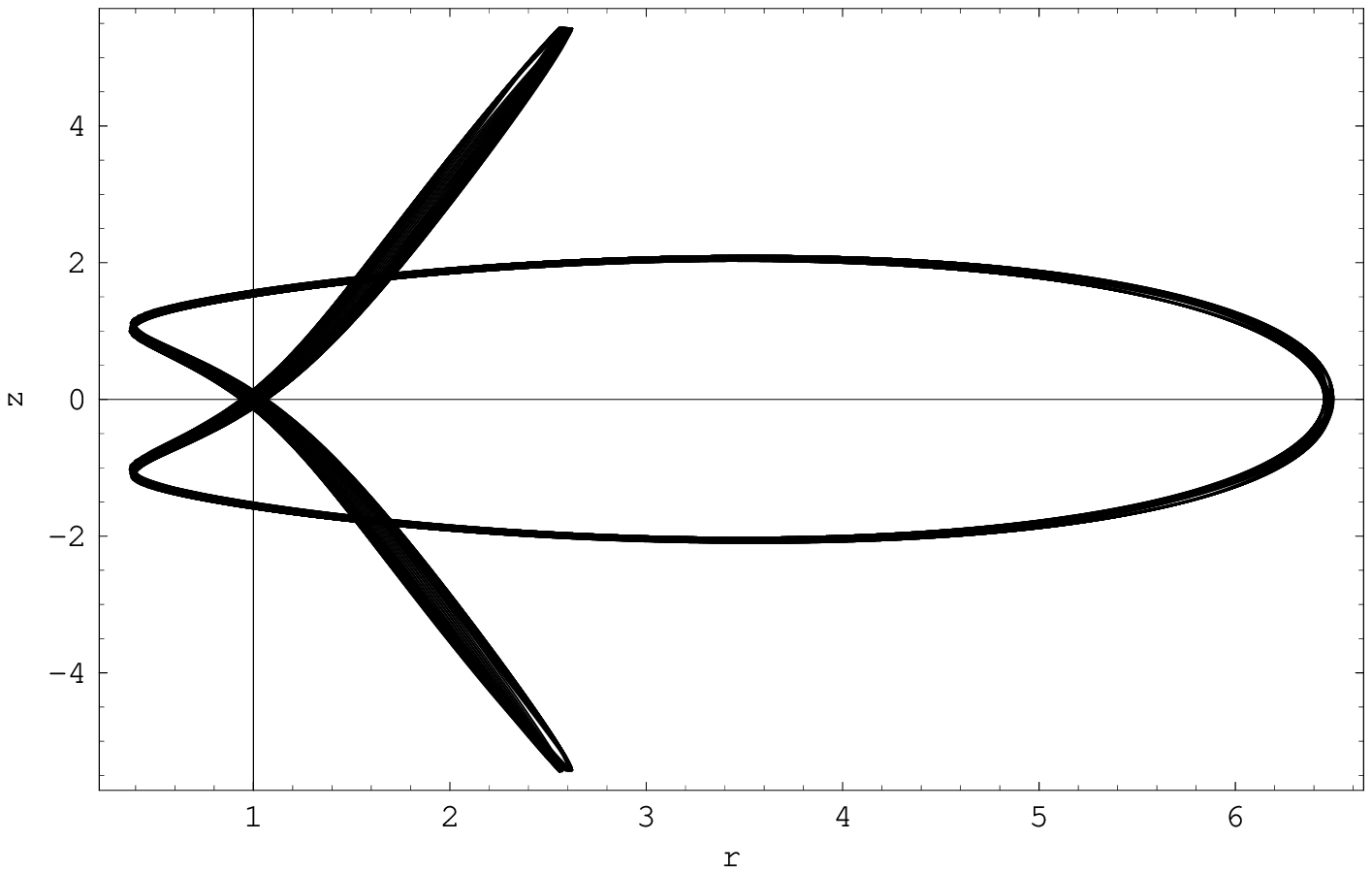}}\hspace{1cm}
                      \rotatebox{0}{\includegraphics*{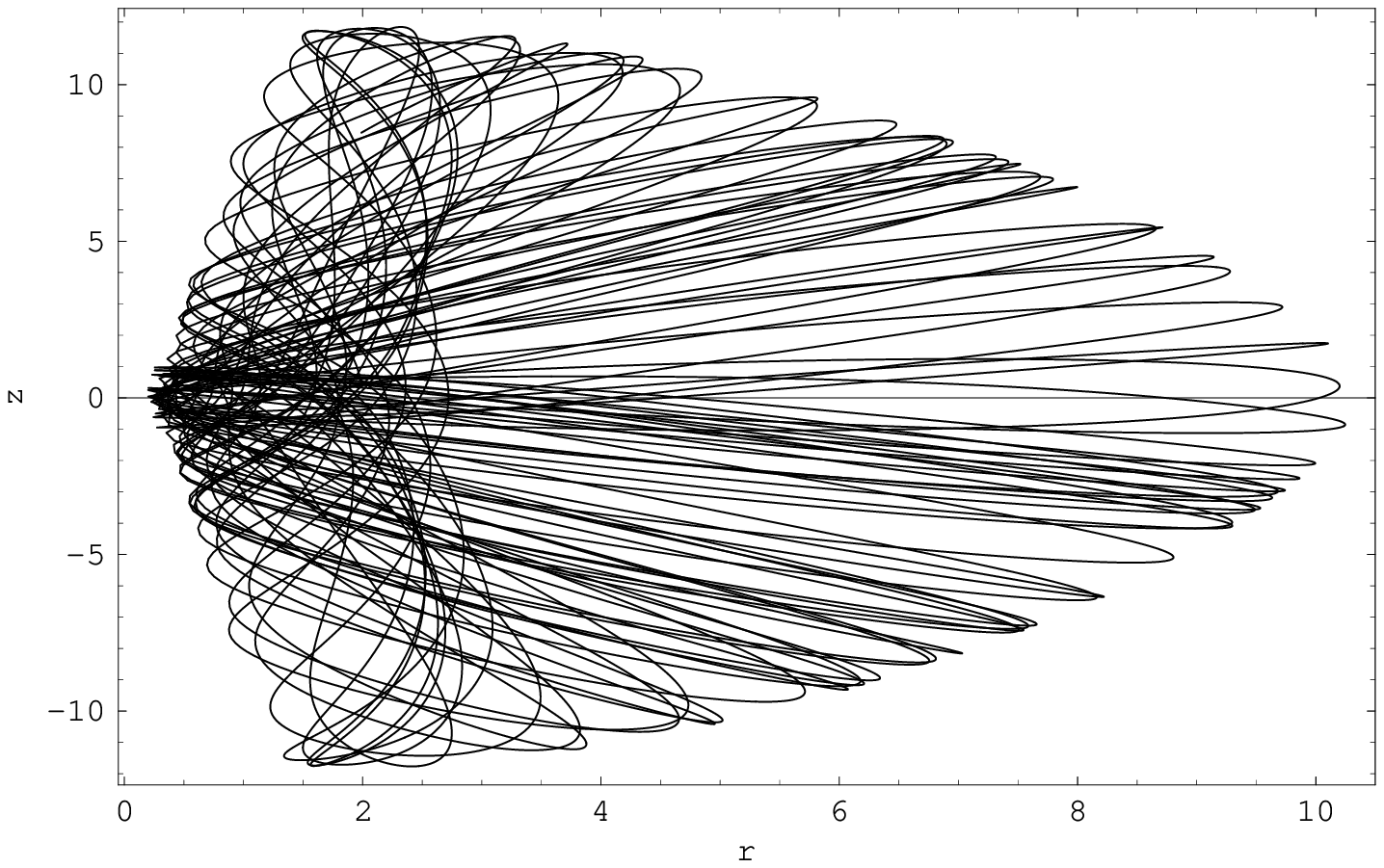}}}
\vskip 0.1cm
\caption{(a-d): Orbits when the system has a prolate halo. (a, \textit{upper left}): an orbit when $\beta = 0.2$ and $c_h = 8.5$. The initial conditions are: $r_0 = 1.3, z_0 = 0, p_{r0} = 6$. (b, \textit{upper right}): a regular orbit when $\beta = 0.3$ and $c_h = 8.5$. The initial conditions are: $r_0 = 5, z_0 = 0, p_{r0} = 0$. (c, \textit{lower left}): a quasi periodic orbit is shown. Here $\beta = 0.6$ and $c_h = 16$. The initial conditions are: $r_0 = 6.5, z_0 = 0, p_{r0} = 0$. (d, \textit{lower right}): a chaotic orbit when $\beta = 0.2$ and $c_h = 8.5$. The initial conditions are: $r_0 = 2, z_0 = 0,
p_{r0} = 0$. The values of all the other parameters are given in text.}
\end{figure*}

Computation of the L.C.E in the case of the prolate halo for different values of $\beta$ and $c_h$, shows that the L.C.E is in the range from about 0.40 to about 0.55, when $c_h = 8.5$ and $0.1 \leq \beta < 0.9$, while it is in the range from about 0.51 to about 0.57, when $\beta = 0.6$ and $8.5 \leq c_h \leq 21$. Figure 10 shows the L.C.E for the chaotic orbit of Fig. 9d.
\begin{figure}
\resizebox{\hsize}{!}{\rotatebox{0}{\includegraphics*{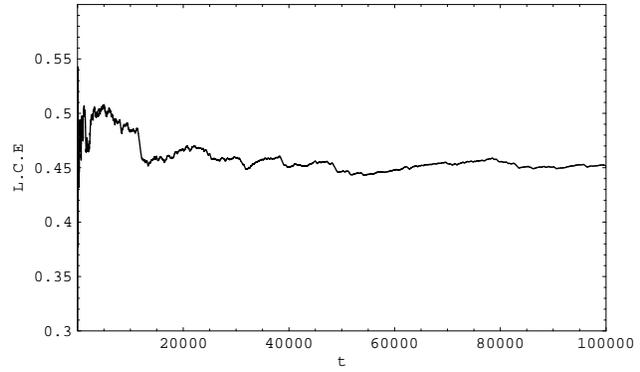}}}
\caption{Evolution of the L.C.E with the time for the chaotic orbit shown in Fig. 9d.}
\end{figure}

Let us now come to see how the percentage of the chaotic regions in the phase plane is connected with the flatness parameter $\beta$. The results are given in Fig. 11 when $0.1 \leq \beta < 2$, that is, for both the prolate and the oblate halo together. There are tree different linear parts in this figure. In the first part, where $0.1 \leq \beta \leq 0.9 A\%$ decreases, in the second part where $0.9 < \beta < 1.0 A\%$ increases very rapidly, while in the third part where $1 < \beta < 2 A\%$ also increases. An explanation for this behavior of the system will be given in the next section.
\begin{figure}
\resizebox{\hsize}{!}{\rotatebox{0}{\includegraphics*{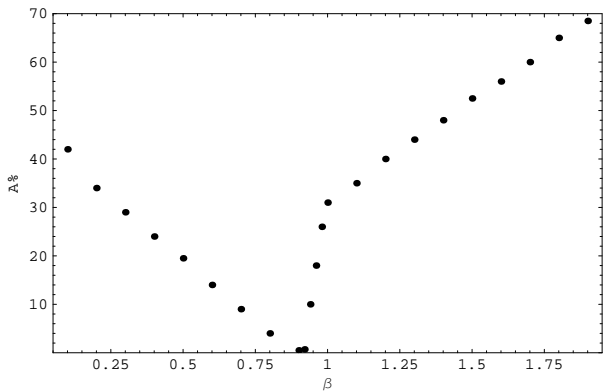}}}
\caption{A plot of the percentage of the area $A\%$ in the phase plane covered by chaotic orbits vs. $\beta$. The value of $c_h$ is 8.5, while the values of all the other parameters are: $\upsilon_0 = 20, \alpha = 3, b = 6, h = 0.2, M_d = 5000, M_n = 400, c_n = 0.25$ and $L_z = 10$.}
\end{figure}

Before closing this section, we would like to present a relationship connecting the critical value of the angular momentum $L_{zc}$ (that is the maximum value of the angular momentum, for which, stars moving near the galactic plane are scattered to the halo displaying chaotic motion, for a given value of $\beta$) and the flattening parameter $\beta$, when all the other parameters are kept constant. The results, which were found numerically, are given in Fig. 12. The value of energy is $E = 600$, while the values of all the other parameters are: $\upsilon_0 = 20, \alpha = 3, \beta = 6, h = 0.2, M_d = 5000, M_n = 400, c_n = 0.25$ and $c_h = 8.5$. We see that the relationship between $\beta$ and $L_{zc}$ is linear. Orbits starting in the upper part of the $\left(\beta, L_{zc}\right)$ plane are regular, while orbits starting in the lower part of this plane including the line are chaotic. An explanation of this behavior will be given in the next section.
\begin{figure}
\resizebox{\hsize}{!}{\rotatebox{0}{\includegraphics*{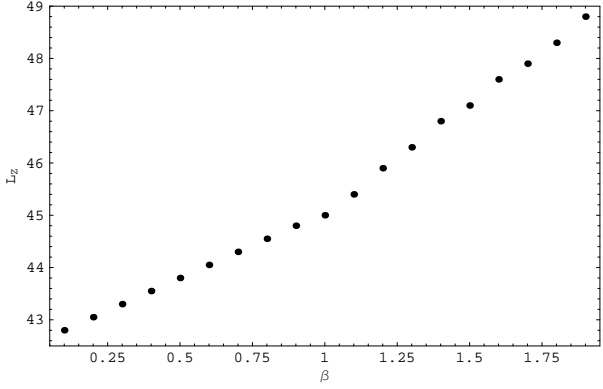}}}
\caption{A plot of the $L_{zc}$ vs. $\beta$ when $c_h =8.5$. The value of $c_h$ is 8.5, while the values of all the other parameters are: $\upsilon_0 = 20, \alpha = 3, b = 6, h = 0.2, M_d = 5000, M_n = 400$ and $c_n = 0.25$.}
\end{figure}

\section{Some semi-theoretical arguments}

In this section we shall present some theoretical arguments together with elementary numerical calculations in order to explain the numerically found relationships given in Figs. 11 and Fig.12. The contribution of the halo component to the chaotic regions observed in the $(r,p_r)$ phase plane of the system comes from two parts. The first part is the $F_{zh}$ force, which is the vertical force of the halo and the second part comes from the asymmetry of the dark halo component. The $F_{zh}$ force is
\begin{flalign}
&F_{zh} = \frac{-\beta \upsilon_0^2 z}{r^2 + \beta z^2 + c_h^2},&
\end{flalign}
while the asymmetry of the dark halo component can be expressed using the ellipticity (see Binney \& Tremaine 2008). The ellipticity of the halo is defined as
\begin{flalign}
&\epsilon = 1 - \frac{b}{a},&
\end{flalign}
where $a$ and $b$ are the major and the minor axis of the biaxial halo respectively. For the prolate halo we find $\epsilon = 1 - \sqrt{\beta}$, while for the oblate halo $\epsilon = 1 - \sqrt{1/\beta}$. Figure 13 shows a plot of the $|F_{zh}|$ vs. $\beta$. As the scattering occurs near the nucleus it must be $r < 1$ and $z < 1$. The particular values of $r$ and $z$ are irrelevant. Here we choose $r = r_0 = 0.2$ and $z = z_0 = 0.1$. We see that $|F_{zh}|$ increases linearly with $\beta$. In
the same figure a plot of $\epsilon$ vs. $\beta$ is given. The value of $\upsilon_0$ is 20, while $c_h = 8.5$. We observe that when $0.1 \leq \beta \leq 0.9$ we have an exponential decrease of $\epsilon$, while at the same time we have a linear increase of the $|F_{zh}|$. The result of these two contrary actions is the observed linear decrease of chaos shown in Fig. 11. When
$0.9 < \beta < 1.0$ the decrease of the ellipticity can be considered linear. But at the same time, we have a considerable linear increase of the $|F_{zh}|$, which has obtained larger values. The outcome is a rapid increase of the chaotic region. When $1 < \beta < 2$, $\epsilon$ increases slowly, while the dominant role is played by the $|F_{zh}|$. As a consequence we see that the chaotic region increase to high values up to about $A\%=70$, when $\beta$ reaches 2.
\begin{figure}
\resizebox{\hsize}{!}{\rotatebox{0}{\includegraphics*{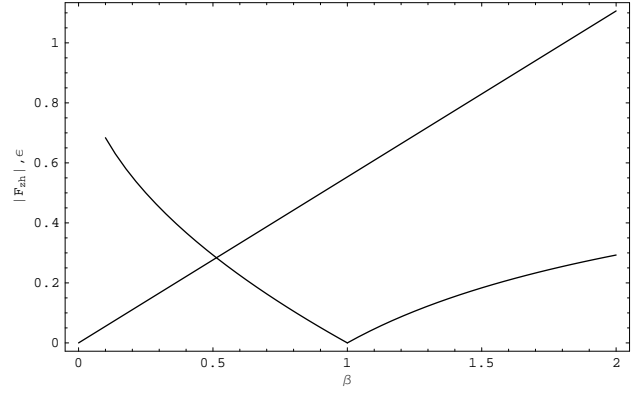}}}
\caption{A plot of $|F_{zh}|$ and $\epsilon$ vs. $\beta$. The values of all the other parameters are given in the text.}
\end{figure}

The exponential decrease of the percentage of the area $A\%$ covered by chaotic orbits in the phase plane can be easily explained by plotting the $|F_{zh}|$ force, given by Eq.(8) vs. $c_h$, while keeping all the other parameters fixed. The results are shown in Fig. 14, where we have taken $\upsilon_0 = 20, \beta = 1.8, r = r_0 = 0.2$ and $z = z_0 = 0.1$. We see that $F_{zh}$ decreases exponentially as $c_h$ increases.
\begin{figure}
\resizebox{\hsize}{!}{\rotatebox{0}{\includegraphics*{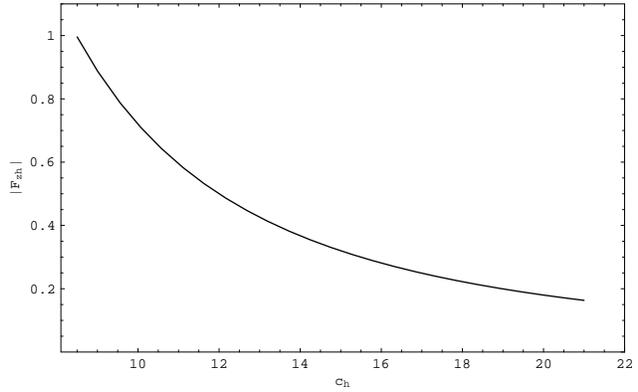}}}
\caption{A plot of $|F_{zh}|$ vs. $c_h$. The values of all the other parameters are given in the text.}
\end{figure}

In order to explain the results shown in Fig. 12 we use an analysis similar to that used in Caranicolas \& Innanen (1991). When the star approaches the dense nucleus its momentum in the $z$ direction changes according to the equation
\begin{flalign}
&m \Delta \upsilon_z = \langle F_{zt} \rangle \Delta t,&
\end{flalign}
where $m$ is the mass of the star, $\Delta t$ is the duration of the encounter and $\langle F_{zt} \rangle$ is the total average $F_z$ force. It was observed that the star's deflection into higher $z$ proceeds in each case cumulatively, a little more with each successive pass by the nucleus and not with a singe ``tragic" encounter. It is assumed, that the star is scattered off the galactic plane after $n > 1$ encounters when the total change in the momentum in the $z$ direction is of order of the tangential velocity $\upsilon_{\phi} = L_{zc}/r$. Thus, we have
\begin{flalign}
&m \sum_{i=0}^n \Delta \upsilon_{iz} \approx \langle F_{zt} \rangle \sum_{i=0}^n \Delta t_i.&
\end{flalign}
If we set $m = 1, \sum_{i=0}^n\Delta \upsilon_{iz}=L_{zc}/r, \sum_{i=0}^n \Delta t_i = T_c$ and
\begin{flalign}
&F_{zt} \approx k_1 + \frac{\beta \upsilon_0^2 |z|}{r^2 + \beta z^2 + c_h^2},&
\end{flalign}
where $k_1$ stands for the constant at a given point and for fixed values of the involved parameters and the vertical force coming from the nucleus and the disk components, in Eq. (11) we find
\begin{flalign}
&\frac{L_{zc}}{r} \approx k_1 + \frac{\beta \upsilon_0^2 |z| T_c}{r^2 + \beta z^2 + c_h^2}.&
\end{flalign}
The star must go close to the nucleus in order to be scattered. In this case we may set $r = r_0 = |z| = c < 1$ in (13) and obtain
\begin{flalign}
&L_{zc} \approx k + \lambda \beta,&
\end{flalign}
where $k$ and $\lambda$ are constants. Relation (14) explains the numerically found relationship between $\beta$ and $L_{zc}$. The different slope in Fig. 12 for the prolate and the oblate halo can be explained because the values of $c$ and $T_c$ are slightly different for the two types of the dark halo component.

\section{Discussion}

In this article, we have tried to study the regular or chaotic character of motion in an active galaxy model with a biaxial dark halo component. It is well known from our previous work in disk galaxies with dense nuclei (see Caranicolas \& Innanen 1991; Caranicolas \& Papadopoulos 2003) that low angular momentum stars moving near the nucleus are scattered off the galactic plane displaying chaotic motion. Furthermore, this procedure is one of the main mechanism that produces chaos in galaxies (see Grosb{\o}l 2002).

On the other hand, earlier work (Caranicolas 1997; Papadopoulos \& Caranicolas 2006) indicates that the role of a spherically symmetric dark halo in galaxies with dense massive nuclei is to reduce the extent of the chaotic regions. It was this motive that drive us to consider a model of an active galaxy with a biaxial halo and study the dynamical effects of the additional component in the behavior of the orbits. In order to keep things simple, we have kept all the parameters of the model constant and studied the behavior of orbits varying only two basic parameters, that is the flattening parameter $\beta$ and the scale length $c_h$ of the halo. In some cases we had to find the critical value of the angular momentum in order to connect it with the flattening parameter $\beta$.

It was found that when a biaxial halo component is present there is a linear relationship between $A\%$ and the flattening parameter $\beta$. On the contrary, the relation between $A\%$ and $c_h$ is not linear but exponential. In both cases the numerically found results were explained using some semi-theoretical arguments. In the same sense we have explained the numerically obtained relationship between $L_{zc}$ and $\beta$.

Computation of the L.C.E shows that the degree of chaos is similar to that found for 3D time-dependent axially symmetric galactic potentials (see Caranicolas \& Papadopoulos 2003), where the L.C.E was about 0.5, while in 2D non axially symmetric potentials (Papadopoulos \& Caranicolas 2006) the L.C.E was found larger, about 1.

The main conclusion of this research is that the presence of a flattened dark halo component in an active disk galaxy has as a result to increase the extent of the chaotic regions observed in the $(r,p_r)$ phase plane. This result is different from the results obtained in earlier papers, where the presence of the spherical halo had as a result the decrease of the chaotic region. It is in our plans, to study the behavior of orbits in active galaxy models, when a triaxial dark halo component will be present, in the near future.

\emph{Acknowledgements}: The authors would like to thank the referee A. Elipe for his useful suggestions and comments.

\end{document}